\documentstyle[12pt,aaspp4]{article}

\def\etal{et\,al.\,}

\begin{document}

\title{Ten Year Review of Queue Scheduling of the Hobby-Eberly Telescope}

\author{
Matthew~Shetrone,\altaffilmark{1}
Mark~E.~Cornell,\altaffilmark{2}
James~R.~Fowler,\altaffilmark{1}
Niall~Gaffney,\altaffilmark{3}
Benjamin~Laws,\altaffilmark{4}
Jeff~Mader,\altaffilmark{5}
Cloud~Mason,\altaffilmark{2}
Stephen~Odewahn,\altaffilmark{1}
Brian~Roman,\altaffilmark{1}
Sergey~Rostopchin,\altaffilmark{1}
Donald~P.~Schneider,\altaffilmark{6}
James~Umbarger,\altaffilmark{2}
and
Amy~Westfall\altaffilmark{7}
}

\altaffiltext{1}{
  McDonald Observatory, University of Texas,
   Fort Davis, TX.
}
\altaffiltext{2}{
  McDonald Observatory and Department of Astronomy,
   University of Texas, Austin, TX.
}
\altaffiltext{3}{
  Space Telescope Science Institute,
   Baltimore, MD.
}
\altaffiltext{4}{
  Population Research Center, University of Texas, Austin, TX.
}
\altaffiltext{5}{
  W. M. Keck Observatory, Kamuela, HI.
}
\altaffiltext{6}{
  Department of Astronomy and Astrophysics,
   Pennsylvania State University,
   University Park, PA.
}
\altaffiltext{7}{
  Information Services, Texas Tech Library      
   Texas Tech University,
   Lubbock, Texas.
}


\begin{abstract}

This paper presents a summary of the first 10 years of operating the Hobby-Eberly Telescope (HET) 
in queue mode.  The scheduling can be quite complex but has worked effectively for obtaining 
the most science possible with this uniquely designed telescope.  The queue must handle dozens 
of separate scientific programs, the involvement of a number of institutions with individual 
Telescope Allocation Committees as well as engineering and instrument commissioning.  
We have continuously revised our queue operations as we have learned from experience.   
The flexibility of the queue and the simultaneous availability of three instruments,
along with a staff trained for all aspects of telescope and instrumentation operation,
have allowed optimum use to be made of variable weather conditions and have 
proven to be  
especially effective at accommodating targets of opportunity and engineering tasks.
In this paper we review the methodology of the HET queue along
with its strengths and weaknesses.

\end{abstract}


\section{Introduction}

The past three decades have witnessed growing interest in moving from the
traditional observatory operation model, in which a single observer (or program)
is assigned a series of dedicated nights to a queue-scheduled mode, where
an observatory team obtains observations on a variety of programs submitted
by investigators who no longer travel to the telescope themselves.
The queue model of observing has been implemented by many major observatories;
some of the early adopters, such as {\it Hubble Space Telescope}, 
{\it ROSAT}, and the
Very Large Array, have become 
standards for comparison.   The great promise of the queue model of
observing is the increase in science productivity of an observatory that 
arises from the flexibility of choosing from a variety of options at a given
time in response to changing conditions and priorities.

This optimization may take many forms,
such as observing the highest ranked targets and programs as assigned by an
allocation committee, 
matching the observing conditions to the constraints of the
observing programs, permitting rapid access for targets of 
opportunity (a particularly difficult challenge
for dedicated night observing),
and allowing for time-constrained observations.
An additional benefit of the queue 
is that experts with the telescope and instrumentation conduct the 
observations; this allows the Principle Investigator to avoid the time and
expense of traveling to remote sites and training on unfamiliar 
systems, and can lead to better quality data obtained by those who are
more experienced with the details of the telescope and instruments.

However, there are disadvantages to implementing queue scheduling.  The 
operational budget of a queue-scheduled telescope, unless it is automated or involves 
instrumentation that is very simple to operate and maintain, 
is considerably larger than 
that of the standard dedicated night model.
There are also widely recognized sociological disadvantages to queue operations, 
which were summarized by Boroson~(1996):
\begin{quote}
"The arguments from the communities who today use these telescopes or their 
predecessors against flexible scheduling include 1) doubts about the 
quality of data taken by observers other than themselves, 
2) the loss of their ability to make discoveries by quick follow-up 
observations, 3) the loss of their overall control of the program 
(the weather lottery turns into the queue lottery), and 4) the loss of 
their ability to be creative and innovative at the telescope."
\end{quote}

A number of ground-based observatories have experimented with and/or
implemented queue observing over the past 10 years; examples of these
efforts can be found in Saha et al.~(2000), Martin et al.~(2002),
Robson~(2002), Kackley et al.~(2004), McArthur et al.~(2004), 
Comeron et al.~(2006), and Puxley \& Jorgensen~(2006).  
In this paper we describe the processes and the lessons learned while
operating the Hobby-Eberly Telescope (HET) for a decade in a queue mode.

The HET is a 9-m class Arecibo-type optical 
telescope located at McDonald Observatory near Fort Davis, Texas
(Ramsey et al.~2007).  The HET was designed for narrow-field spectroscopy
of faint objects; to minimize the cost, the structure has a fixed
zenith angle of~$35^{\circ}$.  The telescope can rotate in azimuth, but
during an observation the telescope is stationary; the sidereal motion
is tracked by a prime-focus instrument package that moves along
the focal surface.
The HET primary consists of 91 identical one-meter hexagonal 
Zerodur segments with a spherical figure; these segments combine to form an
11x10~m hexagonal primary mirror.  The spherical aberration
corrector, carried in the prime focus instrument package, has a pupil
with a diameter of 9.2~m.

The HET can access declinations (decl.) between
$-11^{\circ}$ and~$+71^{\circ}$
($\approx$~70\% of the sky accessible from McDonald), but because of the fixed
altitude of the telescope, the observable area of the sky is a ring centered
at the zenith.  On a given night an object is observable at most twice for
a period of an hour or two (the precise availability depends on the
decl. of the object) as the motion of the celestial sphere causes
the object to enter and leave the ``observing ring" (see Ramsey et al.~2007).  
Figure 1 shows a projection of the ``observing
ring" in hour angle and declination.  Targets at a decl.$ = 30^{\circ}$ 
have two tracks each about 1.2 hr long separated
by 4 hr.  Targets in the north ($decl. > 65^{\circ}$) 
and south ($decl. < -4^{\circ}$) have a single track.

The HET has three facility instruments that are in principle available
at all times: the Marcario Low Resolution Spectrograph (LRS; Hill et al.~1998), mounted
in the prime-focus tracker; the Medium Resolution Spectrograph
(MRS; Ramsey et al.~2003); and the High Resolution Spectrograph
(HRS; Tull~1998).
The latter two instruments are fiber-fed spectrographs located in an area beneath
the telescope structure.

The design of the HET naturally lends itself to the queue mode 
of observing; indeed, it is quite difficult to efficiently operate the
telescope in the standard dedicated night mode.
Because each target has only one or 
two relatively narrow windows of opportunity to be observed on a given night, 
every observation can be viewed
as time-critical.  For example, consider an observing 
project that required one specific object to be observed for a total of
7 hr; this
program could be completed in a single night at a telescope with 
a classical design.   However, the same program would require
more than three nights at the HET.  To effectively execute a single program
on a night at the HET, the program must consist of a
list of targets whose possible HET observing times nicely
mesh with each other (e.g., regularly spaced in right ascension at a given
decl.), but even such a carefully constructed plan would suffer
practical limitations: standards, high-priority objects that 
conflict between east and west tracks,
and the inevitable glitches that occur during a night would
wreak havoc with a tightly packed, time-critical schedule.

Given this situation, queue scheduling of the HET was an integral part of the
HET design, allowing 
the HET to execute temporal projects, targets of opportunity, and surveys 
as part of normal operations.  One additional complication to
the scheduling arises from the governance of the HET; the telescope is
a joint project of five institutions:
the University of Texas at Austin,
the Pennsylvania State University,
Stanford University, the Ludwig Maximilians-Universit\"at M\"unchen,
and the Georg-August-Universit\"at at G\"ottingen.  Each institution has
its own unique
share of the time.  The queue must also keep the partner shares (including
dark and light time) in balance.  

The next three sections of this paper
describe the evolution of operations as experience
revealed shortcomings in the scheduling algorithm.  Section 5
discusses the queue observing in the larger perspective and what works 
well and what does not.   The final section presents future plans for 
the HET queue observing.

\section{Development of HET Queue Scheduling}

The initial concept for HET operations expected that 85\% of the nights 
would be conducted in queue mode and 15\% would be assigned in the classical
dedicated observer mode. 
The staffing proposed for this model was
two full-time Resident Astronomers (RAs) and three full-time Telescope 
Operators (TOs).  The RAs were responsible for target selection, configuration of the instruments, 
execution of the observations, and monitoring of the data quality, while
the TOs were responsible for moving the telescope 
to the science targets, guiding once the target is acquired, and monitoring 
the weather conditions.  The RAs were expected to be active researchers, with
25\% of their time allocated for personal research, access to research and travel 
support, and access to the HET and McDonald Telescopes through the University
of Texas allocation.  The TOs were not expected to be active researchers, but 
have been encouraged to participate (up to 7\% of their time)
in professional development in approved projects that 
interest them.  To provide adequate staffing for
continuous operation of the telescope, the HET facility manager 
was expected to spend half of the time as a RA.

First light of the HET occurred in 1996~December, but the first formally
charged 
queue observations were not taken until 1999~October.  During this time
frame the development of the queue received another level of 
complexity with the addition of the National Optical Astronomical Observatory
(NOAO) to the HET time allocation process.  (The National Science Foundation
partially funded the HRS and MRS; one condition of this support
was that the astronomical community be given partial access to
the HET.  The public time on HET is administered by NOAO.)

Each institution, including NOAO, was responsible for creating its own
HET proposal system, which was denoted "Phase~I"; i.e., each participant
empaneled an independent Telescope Allocation Committee (TAC), and there
was no interaction or coordination between the TACs.  The TACs allocate
time in hours, not nights; this assignment includes the acquisition,
exposure, and detector readout times, as well as
any nonstandard calibration requirements.  Typical
calibrations, such as flat fields and wavelength/photometric calibrations
that can be used by many observations for a given night are
considered part of the observatory overhead and are not charged to individual
programs.  The initial plan adopted by the HET was for the TACs to
distribute their share of the telescope time in three priorities, 
1, 2, and~3, with 1 being the most important programs.
Priority~1 was reserved for targets of opportunity or especially time 
critical observations.  
The expectation was that each TAC would submit a list of programs whose
targets were widely distributed 
across the sky and had an appropriate distribution of required observing
conditions (e.g., transparency, seeing, sky brightness).

The five HET partners adopted the McDonald Observatory trimester schedule
(proposals every 4 months).  At the start of a new
period, all proposals from the previous observing cycle were removed from the
queue.  The TACs do grant long-term status to some programs but the 
Principle Investigator (PI) must
resubmit a new list of targets at the beginning of each trimester.   
The NOAO HET TAC operates on a semester schedule; unobserved
priority~1 and~2 targets remain in the queue until completed, while
priority~3 targets are removed from the queue at the end of each semester.
The NOAO methodology of keeping targets in the queue beyond the end of an
observing period has the advantage of completing high-priority programs 
with minimal effort from the PIs and TACs.  
The five HET partners' methodology of having
the PIs submit new proposals each trimester, with the exception of a few
programs granted long-term status, offers the advantage of keeping the PIs
engaged with the program; reducing their data to the point of being able
to offer progress reports to their TACs.  

The individual PIs are informed of their allocations
by the TACs.  To activate the programs, the PIs submit to the HET RAs
the ``Phase~II" information required to execute the observations:
(1)~the proposal, which includes the science goals of the program
and a clear description of the data acquisition process, (2)~finding charts for
each target, and (3)~an electronic file that contains information
for each requested observation.  This electronic target file consists of a
program number 
(assigned by the TAC), name, coordinates,
$V$-band magnitude of the target, the instrument configuration, 
the number of visits for the target, the total CCD shutter-open time on 
each visit, the number of exposures into which that total CCD 
shutter-open time is to be split, and constraints on image quality, sky brightness,
and transparency.   

Since the TACs are independent, it is inevitable that conflicts would
arise in the accepted programs.  Examples include identical observations
of the same objects, requests for targets of opportunity (in particular,
supernovae and gamma-ray bursts), and a few issues that are produced by the
design of the HET (e.g., if one institution gives very high priority to a
program consisting of a large number of visits to a specific field, then
other targets that are accessible during that time period will rarely be
observed).  There is no ``HET super-TAC" that creates a unified observing
program, but we have developed mechanisms for resolving these conflicts.  

The RAs examine the Phase~II information and identify problems in the
overall program.  In addition to the conflicts mentioned above, occasionally
the combined programs leave a dearth of observations for a specific
sidereal time (a "hole in the queue").  Almost all conflicts have been resolved
by contacting the relevant TACs; a few cases had to be decided by the 
HET Scientist, who is appointed by the HET Board of Directors and whose
charge is to maximize the scientific productivity of the telescope.  The
HET Scientist operates independently of the individual partners and the HET
staff.  The HET Scientist chairs the HET User Committee, which draws its members 
from all HET partner institutions and both serves an advisory role
to the HET Scientist and acts as the conduit of information to the observers
at all partner institutions.

The electronic files submitted by the PIs are collected into a single
database at HET as a single input file for the queue-scheduling
software.
The initial software design for queue observations was inspired by Spike,
which was developed for the {\it Hubble Space Telescope} queue
(Johnston 1991).
The early HET queue
software created a Tcl/Tk graphical user interface (GUI) that 
could present a list of 
targets available at any given time, sorted by various single-selection 
criteria.   In addition, the software creates
a list of the highest priority bright targets and the highest priority dark 
targets for every 15 minute interval for each night.  This basic software
package, denoted "htopx", was used through 2006.

While observing, the RA would frequently run htopx throughout the night
to plan and update the night's observing schedule in real time.  
The calibrations needed for
the night's suite of observations (frequently obtained with more than one
instrument, with occasional multiple configurations per instrument) were
also performed, usually at morning twilight.

At the end of each night, all of the data acquired,
including the calibration files, are transferred to an ftp 
site for individual PIs to retrieve.  Each approved program has a separate
directory to receive those program's observations (only the PI has
access to the data for a given program).  
The calibrations are placed in a directory that all PIs can access.
When the data are in place, the PI
is notified via e-mail that new data are available; this message also
contains any concerns that the RA has regarding the quality of the
data or presents any difficulties that were encountered in the observation.
This system
allows PIs to provide almost immediate feedback about the quality
of the observations or to suggest adjustments to the observing 
strategy for future observations.

\section{Early Operations for HET Queue and Observing Support}

One of the first advantages of queue
scheduling became clear in the earliest days
of HET operations: engineering time and instrument 
commissioning were much more effectively combined with science observations
than with traditionally scheduled 
telescopes.   When an instrument commissioning run had to be rescheduled
(often on short notice), it was trivial to shift the telescope back to
science operations.
Similarly, when a problem arose that required an
engineering effort, it was easy to reschedule the science program.  Thus,
an individual astronomer would not bear the entire cost of the loss of 
telescope availability.
In addition, the local and visiting 
engineering staff could halt their work for a brief time,
when feasible, to allow an observation to be obtained of a high-priority target.
Queue scheduling had the added benefit of allowing the engineers to release the
telescope to the RA when their work was completed ahead of schedule or
had reached an impasse.

The initial plan of having two full-time RAs and the duties of a half-time 
RA filled by the facility manager proved unfeasible for two reasons:
(1)~the facility manager was required 
to spend all of his time with the daytime support staff to keep the facility 
running smoothly,
and (2)~there were never any requests 
from PIs for dedicated time.  The HET immediately became
a 100\% queue-scheduled telescope by default, and hence would be 
understaffed.  To address this
staffing shortfall, a third full-time RA position was created.

Target conflicts between the various TACs or PIs have been very rare; 
since 1999, there have been requesting conflicts with two 
gamma-ray bursts, four supernovae and two extra-solar planets.
The policy that we 
developed calls for sharing data among the Target of Opportunity (ToO) 
requests and for coordinating the 
instrument setups.  For non-ToO targets our policy has been to observe
each request separately according to the queue scheduling algorithm 
and to distribute the data separately. 

During early operations the overhead times charged to each 
PI's program 
were found to be highly
unpredictable, with some visits taking a factor of 2 longer than the average length 
due to technical problems with the telescope, including variable image 
quality and mirror alignment issues.  To give the PIs a sense of 
predictability during this period of early operations we set a standardized 
overhead of 10 minutes per requested visit until the actual overheads 
could be more predictable, which occurred once full science operations began.

Despite the e-mails that the PIs received each night, the PIs found that they  
needed a Web-based interface to 
monitor the progress of their programs.   
To create this page it was necessary to access the night 
reports (where the log of observed spectra were kept), the electronic 
Phase II (where the log of the completed observations were kept), and a 
file containing the TAC allocations for each program.  Had this demand been 
anticipated these various data sets would have been stored in a single database.
The resulting Web page shows the 
observations that have been attempted, the targets in the queue, along with 
their status 
(completed or active), and the amount of TAC telescope allocation used for each program.   

Target selection by the RAs at the telescope was not found to adequately reflect the 
TACs' wishes.  Most of the time, the RA would be forced to choose from a very wide
range of targets, all with the same TAC priority.  
The three-priority system did not offer enough dynamic range to represent
the TACs' scientific ranking. 
In addition, there were times when significantly poor seeing or transparency 
limited the available targets, but some types of science could still have been conducted. 
The TACs did not want their institutions to be charged for the long setup times 
(due to the 
difficult conditions) and extended exposures (again due to bad 
conditions) for these normally easy targets.  To address these problems we 
developed a new priority scheme, 0--4:   

Priority 0. -- Time is allocated for 
targets of opportunity or very time-critical observations.  Up to 25\% 
of the priority 1 time can be assigned to priority 0.

Priority 1. --
Targets would constitute one-third of the expected partner's observing time during 
the period including average weather losses and time lost due to engineering 
and technical problems.

Priority 2. -- Targets would make up the second third of the weather-corrected
partner share. 

Priority 3. -- Time would make up the final third of the 
weather-corrected partner share. 

Priority 4. -- Time is unlimited, but the 
expectation is that this time would only be used when normal operations 
could not be conducted (e.g. moderate cloud coverage or very poor seeing).  
In addition, priority 
4 time would not have any charged overhead and would only be charged at 
half the nominal conditions exposure time.   

The reason for only charging
half the nominal conditions exposure time for priority 4 (P4) targets
is to give the PI some advantage
and incentive to work with data that are acquired under sub-par conditions,
and to compensate the PI for the extra effort that can be required to reduce
such data.

Despite some of the problems mentioned above, some aspects of HET's queue-scheduled 
observing program were immediately successful (see Ramsey et al. 1998).  
The RAs were directly involved in informing the PIs of data delivery 
(and providing comments if useful) and they rapidly responded 
to comments from the PIs.   This highly personalized attention given to the 
PIs required the RAs to make temporary notes in the PI's 
Phase II materials and to communicate with the other resident astronomers about 
these changes.   This effectively added a customer service aspect to the 
job description of the resident astronomer.  The TACs and PIs overwhelmingly 
approved of this aspect of the HET's performance.

To better serve the PIs, several new fields were added to the Phase II files. 
The first was an option for the PI to request radial velocity, spectrophotometric,
telluric or other types of standards.  The second addition allowed the PI to 
request a standard set of calibrations or to make a request for specific 
calibrations, such as a wavelength comparison spectrum immediately after 
their visit.  We also created a flag that would allow the PI to group two
or more observations together, such as a specific standard to be observed
directly after their main target.
The final change to the electronic Phase II specifications
was the addition of two synoptic-specific 
fields: one to specify the frequency of the visits, and the second  
to specify when the visits should occur, using a date string.  The latter allows for 
specifying specific visit dates and/or ranges of dates in which to 
make visits.  A complete list of the entire Phase II format can be found 
on the HET Observing Support Web pages.\footnote{See http://het.as.utexas.edu/HET/hetweb/}

\section{Full Operations with the HET Queue and Observing Support}

Under full science operations, the typical queue contains 1500 entries, with
each entry describing from one to 20 requested visits.  The average
requested number of visits is 1.9.  Typical requested visit times are 
1300 s for LRS and 600 s for HRS (the HRS visits are
dominated by planet-search proposals, which involve a large number of short
visits spread out over the trimester).   We typically have 45 programs at
the beginning of the trimester.  Figure 2 exhibits the cumulative distribution 
of TAC allocations for each priority.   In 2006 the total shutter-open time made
up 50.3\% of the clear science time.  The remaining 49.7\% includes target overheads, 
mirror alignment, science calibrations and time lost due to problems.  In
2006, we lost 31.9\% of the nights to bad weather and spent 3.2\% of the
nights conducting scheduled engineering tests.

Our TACs have found that allocating
a blend of priority classes to a single program allows the more critical 
targets to be observed and gives some flexibility to the PIs.   The projects
that have successfully made use of the lowest priority time, P4, have been
programs that submit a large number of targets distributed widely over
the sky, with exposure times that are short ($<600$ s).
When the number of submitted 
programs is compared with partner share, we find a rough average of one program
for each 2\% share of telescope allocation.   The smallest HET partners have 
roughly 6\% share in the observing time.  During a trimester, we typically also 
receive two to five new
proposals from the TACs.  These are allocated from time that the TACs hold back 
for unforeseen exciting science, such as ToO programs.   

The expectation for 
the partners was that they would, on average, share equally in all observing
conditions.  For some of the partners, this has not been the case.  For
example, in 2006 two of the smaller HET partners had less than 1\% of their 
total submitted targets in bright time (sky brightness V$< 19.5$).  The HET
Board of Directors and HET User Committee do not have any formal mechanism or
policy to deal with this inequity.
By the end of the trimester, bright time is 
largely dominated by P4 targets submitted by the two larger HET partners.  

When an observation is completed, the RA attaches a flag to the night report
entry to indicate if the observation is charged or uncharged.  The charged
category is further broken down into acceptable and borderline.
In 2006, the borderline observations made up 5.3\% of the charged 
observations.   The PIs have the option of requesting an observation be repeated;
usually these are borderline observations that the PI believes are of too poor a
quality to use for their analysis.  In 2006, PIs rejected a total of 
3.1\% of the science-collecting operating hours.  
From 1999 to 2006, there has been
only a single case of a disagreement between an RA and a PI as to whether
an observation should be reobserved at no cost to the PI.

Once the HET entered full science operations it became clear that the 
HET facility manager could not effectively monitor the night-time science 
operations without assistance.   A position was created to manage the 
science operations and supervise the science operations team.  The science 
operations supervisor reports to the facility manager, but also interacts 
with the TAC chairs and the HET scientist.   The science operations 
supervisor maintains metrics on science operations, approves procedures for 
night time operation, monitors the nightly operations, and makes night-to-night 
decisions about the operation of the queue, including balancing 
engineering time with science operations.   The science operations 
supervisor also creates a monthly report for the TACs.  This report describes 
the status of the observing programs and that of the facility, which 
allows the TACs to make changes to the priorities to ensure that 
their most important programs are completed.

Under the three-TO/three-RA staffing plan, 
normal attrition left the HET partially crippled.   The re-hire and 
training process takes approximately 6 months, and operations with just 
two RAs or TOs led to observing inefficiencies, 
decreased morale, and an increased attrition rate.  Just as spares for 
critical components are required for the telescope, we required a plan for 
replacement of critical staff.   The science operations supervisor would 
be required to act as a part-time RA during normal 
operations and to fill in as a full-time RA during any 
hiring and training periods.   A fourth TO was hired, leaving night operations somewhat over-staffed under normal conditions, so extra duties were added to 
the TO position.  These new duties included working in the 
afternoon with the day staff to assist in telescope maintenance and 
engineering operations, and working before sunset to prepare the telescope 
for operations.   A summary of the HET science operations staff is
given in Table~1. The HET scientist is also included in the table, even
though that position is independent of HET operations. The HET has a 
total of eight full time equivalents (FTEs) dedicated to science operations.  
This is 40\% of the 
total FTE compliment for the HET, and is 27\% of the total HET budget.

One of the obstacles to program completion was found to be PIs who 
submit programs with unrealistic observing requests (e.g., extreme image
quality constraints, expectation of a large number of visits).   
A Phase I tool was developed to allow the PI to get a 
sense of which targets and observing programs have a reasonable expectation of
being completed.
Some TACs require that each PI explain the feasibility 
of completing the observing program in the allocation period, based on results 
from this Phase I tool.  Although this tool 
allows a PI to determine if the program is feasible, it does not show 
conflicts that might arise with other programs in the queue.   The most 
common form of conflict is overly subscribed portions of the sky;  
the north Galactic pole and the Coma cluster of galaxies are typical examples.   
A first attempt 
to understand these conflicts is made by breaking down the queue into 
histograms based on the sidereal time (ST) for each requested visit 
and comparing the expected number of visits that may be completed at each 
ST bin.  Figure 1 shows that for targets with declinations between 
$-3^{\circ}$ and $+64^{\circ}$,
there are two available windows 
in which they may be observed (east and west), and that these tracks are
often separated by several hours.   However, during any given observing period 
(trimester), some targets may only have a single track available, since the east or 
west tracks may not be observable because of temporal conflicts with morning or 
evening twilight.

Figure 3 exhibits the histogram for the 2006-1 period for priority~0--3
dark time (sky brightness constraint of V $>=$ 20.6) targets and includes all visits (east
and west).    
The unfilled histograms are the requested dark time 
visits in the queue, and the filled histogram are the 
completed visits.  A target at $decl. = 30^{\circ}$ with a single 
requested visit will appear
twice in the figure, once for the east track and once for the west track.  Similarly,
completed visits in Figure 3 are double counted for targets with both east and west
tracks.  The solid line represents our estimate for target 
completion at each ST, based on the Phase I program completion expectation tool 
that is mentioned above and found on the HET Observing Support Web pages.   
The completed
visits exceed the estimated visits between ST 21 and 0 because these are targets
with double-counted visits; e.g., the east track falls during the day, but the target
was observed in the west track.   
At the beginning of the trimester, the RAs report to the TACs 
any programs that are in jeopardy of not being completed because of target 
density on the sky, e.g., the large number of requested visits over
the number expected to be completed at ST $=$ 10 was largely due to one 
program.  The TACs can report this information to the PIs, or they can
take any appropriate steps to modify the queue.    The RAs can also attempt
to exceed the estimated completion rate for an overly dense region by systematically 
observing targets with two available tracks
at STs that are less densely populated, such as STs between 6 and 8.  This was
done in ST bins  $ 10$ and $ 14$.

While the setup times during full science operations were more predictable than
during early operations, the setup times charged to PIs were found to  
be larger, on occasion, than the assumed setup times adopted in the Phase I proposal, which can
make program completion in the allocated TAC time difficult.  In order to ensure 
that the program could be completed, and to give the PIs 
an element of predictability, a cap is placed on the amount of overhead time
charged to any requested visit.  The overhead cap is instrument dependent.

The choice of targets to observe during night operations is made by
the RA, who bases the selection on a balance of object 
availability (how many more visits can be made to the target during the current trimester), 
TAC priority, synoptic constraints, 
and current observing conditions.  To assist the RA  
with these choices 
and to add a level of standardization, a modified-priority scoring system
was created. 
A set of modifiers are added to the TAC's priority, and the results are sorted 
to aid in target selection.
The modifiers are listed 
in Table 2, along with the maximum magnitude of the modifier.   For
targets that have synoptic constraints,
the PI can give either a flexible or a fixed range;
the synoptic modifier given in Table 2 is for the flexible range. 
The modified-priority system is tuned such that:

\begin{enumerate}
\item The total modifiers on a priority~4 target never allow it to outrank 
a priority~3 target. 
\item The synoptic modifiers on a priority~2 synoptic target can 
allow it to outrank a priority~1 target.
\item The object availability modifier on a priority~2 target can 
allow it to outrank a priority~1 target.
\item The total modifiers on a priority~1 target can combine constructively to 
allow it to outrank a priority~0 target.
\item The total modifiers on a priority~3 target can combine constructively to 
allow it to outrank a priority~1 target.
\end{enumerate}

The modified priority system was created and implemented by the RAs
but it is continuously evaluated by the HET User Committee.   The 
system allows for the creation of an initial observing plan for each
night by selecting the targets with the highest modified 
priorities and then filling out the schedule with lower modified-priority 
targets.  The modified-priority algorithm creates a suggested plan 
for the RA, and it provides some predictability
as to the order in which the targets will be observed, which is useful
for planning instrument changes.  However, 
as the observing conditions change, the RA is still required 
to make critical decisions about which targets to observe.  

Program completion is one of our critical metrics.
For over-subscribed or perfectly subscribed programs (programs that have
enough targets to be completed in the TAC-allocated time), the program completion
percentage is calculated from the amount of TAC-allocated time completed at
each priority.  For under-subscribed
programs (not enough targets to fill their TAC-allocated time), the program
completion percentage is calculated from the completed versus
requested shutter-open time at each priority.
Figure 4 exhibits the median completion rates for the individual 
programs from 2004 through 
2006.   The modified-priority system was implemented at the beginning
of 2005, and further modifications were made in 2006 to fine-tune the 
algorithm.   From this plot, we have concluded that our completion
rates for the highest priority targets (those with the lowest priority number)
 are not driven by the algorithm, 
but by the weather and the feasibility of the program.   The lower
priority targets have a better completion rate under the modified-priority
methodology than under the subjective, RA-specific nightly
decisions.

\section{Review and Status:  Summary of what does and does not works well}

A subjective analysis of the HET science return reveals that we have been most 
successful by concentrating on target-of-opportunity surveys (e.g.,
Frieman et al. 2007), synoptic programs (e.g., McArthur \etal 2004; Kaspi
\etal 2007),
and surveys with a 
wide distribution on the sky (e.g., Sowards-Emmerd et al. 2005).  For
more information about completed science programs see Ramsey (2005) 
and Ramsey \etal (2007).
Most of the incomplete programs 
have been pencil-beam surveys (e.g. many 
visits to a single target, or many targets in a single, small region of the sky). 
Most of the failed programs have failed due to technical problems at the 
telescope and are not 
due to problems with the HET queue-scheduling methods.
While the efficiency benefits of queue scheduling under full science 
operations are documented in the literature (see references quoted in the
introduction), we found 
that in early operations the benefits were even more substantial.  The ability 
to begin or end an engineering effort to attack a subtle or intermittent 
problem without disrupting a specific PI's allocated time, or the ability to 
pause engineering for a very high priority science target made the first 
years of science operations more productive.   Even today, 
planning an engineering run is far easier on a queue-scheduled telescope 
than on the traditionally scheduled telescopes at McDonald Observatory.

The basic lesson learned from the Hobby-Eberly Telescope queue-scheduling 
effort is that a customer-oriented observing effort can be highly successful. 
While the HET would have benefited from greater software development early on, 
the evolutionary manner in which the software and observing styles developed 
has allowed us to understand the needs and improve service to the institutional partners and PIs. 
Since it is the PIs who ultimately will be analyzing the data, making an 
effort to involve them as active participants of the process is critical.  
The HET does this in several ways:

\begin{enumerate}
\item The PIs are allowed and encouraged to e-mail the RAs
of the Hobby-Eberly telescope during their Phase I and 
Phase II planning.
\item The PIs have access to their data on a nightly basis.
\item The PIs can make some changes to their Phase II information after submission.
\item The PI can request that observations that may have been compromised 
by the weather or improperly observed be repeated.
\end{enumerate}

These steps to involve the PI address many (but not all) of 
the concerns that most PIs have about queue scheduling as 
summarized by Boroson~(1996) and listed in the introduction, but only if the 
PIs are encouraged to actually examine the data and provide timely feedback to 
the RAs.   In addition, we have allowed our software tools 
to evolve as the PIs and TACs request new features or 
interfaces.  This necessary evolution arises from a customer-oriented model 
for the science operations.   To further our efforts to improve
customer service, we created a survey to be filled out by the PIs
after Phase II and again half way through the trimester.  Our response rate
for the Phase II survey was 22\%, while that for the 
mid-trimester survey was 28\%, out of 32 PIs.   The results of the survey 
can be summarized as follows: 56\% of the responses had positive comments 
and no constructive criticism,   19\% had constructive comments on our Web 
documentation, 19\% had constructive comments on our Phase II
process, and 6\% had constructive comments on the instrumentation.   While the
feedback was moderately useful, we believe the low response rate did not warrant
a survey be repeated every trimester, since the vast majority of our 
PIs are the same from trimester
to trimester.

One unique feature adopted by the HET is our poor-conditions priority 4 policy.
This has been tremendously useful and successful.   Not only have P4
targets been observed under poor conditions, but the short-exposure targets
are also useful for filling in between high-priority targets on nights with good
weather.  Recall that the HET must wait for specific targets to enter 
the observing annulus on the sky to be observed, so having an abundance of
short-exposure targets in the queue to fill in the "dead-time" leads to 
higher scientific productivity.   In 2006, 35.8\% of accepted CCD 
shutter-open time was for P4 targets.   Even for a conventional telescope, 
there will still be reasons to have a large pool of short-exposure
targets to be observed in poor weather conditions or to fill
in between high-priority time-critical targets.   Without some incentive, 
such as our policy of half-charge for the nominal
shutter-open time and no overhead charges, the PIs and TACs would have reduced 
incentive to submit targets for poor observing conditions, and the 
queue would be less efficient and the telescope less productive.

Another important lesson that we have learned from running fully queue-scheduled 
operations at the HET is the need to have a plan for 
replacement of critical staff.  Specifically, we must be able to cope 
with the normal attrition of TOs and RAs.
This plan can be as simple as identifying who will take up 
the work load until new staff are hired (as is the case for our RAs
and the night operations supervisor) or the more extreme 
measure of having an extra person in the rotation so that the observatory 
is never under-staffed (as is the case for our telescope operators). 
With the increasing costs of telescopes, the expense of having an additional 
telescope operator and resident astronomer is small but crucial.

While the staff require considerable training to properly operate a queue, 
we have found 
that the TACs and PIs also need training to properly 
populate the queue.  The PIs must have a good working knowledge of how the 
telescope and instrumentation operate and what their capabilities are.  
One of the 
areas in which the HET operations staff have been less successful is in maintaining
a living repository of knowledge of how the HET and 
its instruments are performing.   Without this information, the PIs have 
relied on communication with each other to determine if projects are feasible.   
This problem could be solved with one more FTE.  This additional person could
take on the documentation of the facility's capabilities as a principle
duty, or else the person could be designated as another RA and then have the
documentation duties split among the RAs.

Just as educating the PIs is critical, the TACs must be staffed with people
knowledgeable about the telescope's capabilities in order to
determine if the
science proposals have merit.  In the case of the partner institutions this
does not seem to be a problem, since most of the TAC members are or have been
PIs.  There have been some instances in which programs approved by non-HET partners
have not been well suited for the HET; while technically feasible, such
programs have a low expectation of completion.  
For future queue-scheduled telescopes, we recommend that the observatory staff 
participate in the time-allocation 
process during early operations until a significant 
fraction of the TAC members are familiar with the facility.  This participation could
take the form of nonvoting members or as an external review of project
feasibility.
We have found that continued interaction with the TACs or a TAC
representative is the best approach to maximize each institution's science
productivity.  Most of our interaction has been
in the form of monthly reports from which the TACs can follow the progress
of program completion, the institutional shares and the instrument and
facility status.   In addition, the TACs' policy of retaining a small percentage
of their telescope allocation to be
distributed later during the trimester to cover unforeseen opportunities has
frequently yielded significant science results.

One of the major ongoing problems we have had in completing programs has 
come from an unforeseen impact of adding new features and improvements to our 
instrumentation. One of the reasons that HET works well is that the 
instrument complement and design allows HET to rapidly change between 
instruments to take advantage of changes in the observing conditions and 
to complete the most difficult science that the observing conditions allow. 
As more features have been added to the HET instrument complement, 
one of the greatest challenges to queue scheduling has been balancing
requests for mutually exclusive
instrument configurations.  The HET has one instrument, the LRS, that
cannot have all modes available all the time.  Four grisms are used for
LRS observations, but the instrument can hold only two on a given night
(grism changes are limited to daytime operations).  One of the slots
is devoted to the most requested grism (g1).  The demand for the remaining
dispersers (g2, g3, and e2) is quite unequal, with g2 being the most popular
and e2 rarely being requested.  Figure 5 displays the completion rates for
priority~0--3 LRS programs in 2005 and 2006, divided into three groups:
g1, g2, and g3$+$e2.  The median completion rates are roughly
equal for the high-priority programs, largely due to the 
roughly one to four instrument changes per dark run.  For the lower priority
time (larger priority numbers), the completion rates for the less subscribed
setups are considerably lower.  This problem was particularly complex 
when one of the partners with only a small share requested an instrument 
configuration that cannot be mounted at the same time as one of the more 
popular configurations.  The continuing desire to upgrade to specialized 
instrumentation must be balanced against the impact that that upgrade 
would have on the flexibility of the queue to respond to different 
observing conditions and configurations.

Any new collaboration on a queue scheduled telescope should have a well-designed
plan for dealing with conflicts between the competing target requests
and for dealing with partners who wish to specialize in one specific part
of the sky or one specific observing style.  This last issue can manifest
in requests for all dark time or in an institution requesting only one 
specialty instrument.   This situation can lead to the awkward problem of
a partner not receiving their full allocation in an observing period which 
begs the question: what techniques, if any, 
can be employed to rectify the situation?
At the HET we have encouraged any partner who has fallen significantly behind
their partner share to submit more targets and have given those targets extra
emphasis through the modified-priority algorithm.

\section{Next Generation of Queue Scheduling}

Our work with HET and its queue methodology is an ongoing program;  
most of our operations efforts are now going into improving the PI, TAC,
and RA software interfaces.   We wish to automate many of the 
features that are currently performed manually.  For example, 
checking the status of a program with respect to its TAC allocation
and determining whether the program has used its allocation requires one 
to manually place on hold any
remaining targets for that program.  Automation will not only reduce 
the RA work-load, but will also increase reliability by removing the human 
element from the loop.  
Some examples of these changes are:

\begin{enumerate}
\item Automate the checking of program completion.
\item Allow a PI to add or remove targets directly to or from the queue.
\item Allow the TACs to create programs and allocate time directly to the queue.
\item Create an automated list of calibrations and standards to be taken for each target.
\end{enumerate}

One of the strengths of the HET has been the fast response 
to PI requests, and we wish to build on this strength.  Current requests
by PIs for changes to their programs
are not common outside of the synoptic programs.
Medium-term planning has been more sensitive to the site's variable
weather than to changes to PIs' programs, and we anticipate that allowing
the PIs greater control over their programs will not change this.  

These changes will require a transition from the current HTML-based night reports and 
Tcl/Tk$+$tab-delimited text file to an integrated database.   Our choice for 
this upgrade is a 
MySQL database with a variety of interfaces, including HTML, PHP, JavaScript, and
AJAX technologies.

The current working prototype uses PHP sessions to authenticate users, then allows 
for flexible, individual and graduated access to program and operational data. 
Permissions can be granted according to the status of active programs or membership 
on allocation committees. Access to Web applications to plan, administer, and 
evaluate programs, and to interact with the queue can also be customized. It is 
envisioned that a well-designed suite of tools that provides timely and accurate 
information on operational conditions, including queue activity, for PIs, TACs and 
observatory staff has the potential not merely to improve efficiency and 
productivity through automation, but to give all users the power to leverage the 
advantages of the service model to refine their own programs in a dynamic 
queued environment.  

\acknowledgments

We are deeply indebted to the 
night staff of the HET over the last 10 years: Grant Hill, 
Ben Rhoads, Teddy George, Gabrelle Saurage, Frank Deglman, 
Mike Soukup, Michelle Graver, Martin Villarreal, John Caldwell, 
Vicki Riley, Chevo Terrazas, and Heinz Edelmann.   We would 
also like to thank the day staff in West Texas and the 
Austin engineering 
staff for all of their efforts on the HET.  
We would like to thank Larry Ramsey, Rob Robinson,
Roger Romani, and the HET User Committee for their 
continuing guidance.  Support for 
A. W. was under REU program grant NSF AST-0243745.
The Hobby-Eberly Telescope is a joint project of 
the University of Texas at Austin,
the Pennsylvania State University,
Stanford University, the Ludwig Maximilians-Universit\"at M\"unchen,
and the Georg-August-Universit\"at at G\"ottingen.  
The HET is named in honor of its principal benefactors, 
William P. Hobby and Robert E. Eberly.

\clearpage

\begin{deluxetable}{lr}
\tablecaption{Science Operations FTEs}
\tablehead{
\colhead{Position}&
\colhead{FTE}}
\startdata

Science Operations Supervisor & 0.5 \\
Resident Astronomers & 3.5 \\
Telescope Operators & 4.0 \\
HET Scientist & 0.1 \\

\enddata
\tablecomments{The HET Scientist is not formally a part of the 
night operations team.}
\end{deluxetable}

\clearpage

\begin{deluxetable}{lrc}
\tablecaption{Priority Modifiers}
\tablehead{
\colhead{ Modifier   }&
\colhead{ Algorithm   }&
\colhead{Maximum } \\
\colhead{  }&
\colhead{   }&
\colhead{Magnitude } }
\startdata

Priority 0                             & If P0 then  -1  &$-1.0$ \\
Priority 4                             & If P4 then  +2  &$+2.0$ \\
Object availability                    & Log[($N_{\rm visits}-N_{\rm request})/(f_{\rm min}*N_{\rm setups})]$  &$+2.4$ \\
Object completeness                    & $  0.6*(N_{\rm request} - N_{\rm done})/N_{\rm request}$ &$+0.6$ \\
Partner share                          & $  4*(F_{\rm current} - F_{\rm HET})/F_{\rm HET}$ & $-1.5$ \\
Synoptic modifier & If Date $> $Date$_{\rm max}$ then & \\
 &  $-0.6-0.15*$log[(Date$-$Date$_{\rm max})/(f_{\rm max}-f_{\rm min})]$ &$-1.0$ \\
 &  else $-0.6*($Date$-$Date$_{\rm max})/(f_{\rm max}-f_{\rm min})$ &  \\

\enddata
\tablecomments{\\
$N_{\rm visits}$ is the number of visits left in the observing 
period, including any restrictions imposed by firm synoptic deadlines. \\
$N_{\rm request}$ is the number of visits requested by the PI. \\
$f_{\rm min}$ is the minimum length the PI prefers between visits. \\
$f_{\rm max}$ is the maximum length the PI prefers between visits. \\
$N_{\rm setups}$ is the number of exclusive setups that compete with the requested setup. \\
$N_{\rm done}$ is the number of visits completed. \\
$F_{\rm current}$ is the actual cumulative fractional share the partner has. \\
$F_{\rm HET}$ is the cumulative fractional share the partner should have. \\
Date is the current decimal date. \\
Date$_{\rm max}$ is the last day in the preferred observing window.}
\end{deluxetable}

\clearpage

\begin{figure}
\epsscale{0.8}
\plotone{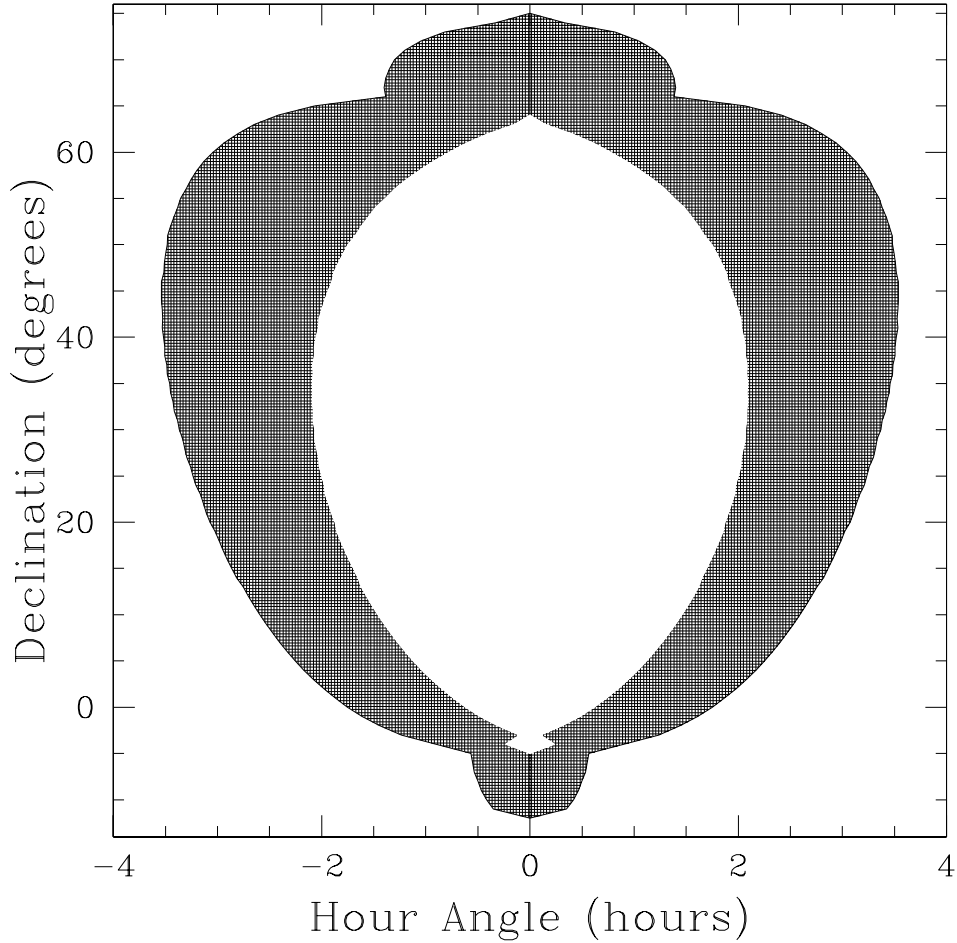}
\caption{Annulus observable by the HET,
translated
to hour angle and declination.  The shaded region is the 
area observable by the HET.  Targets with  $decl < -10$ or 
$decl > +70$ are not observable by the HET.  The longest tracks
for the HET are at $dec = +63$ and are 170 minutes long.
\label{fig1}}
\end{figure}

\clearpage

\begin{figure}
\epsscale{0.8}
\plotone{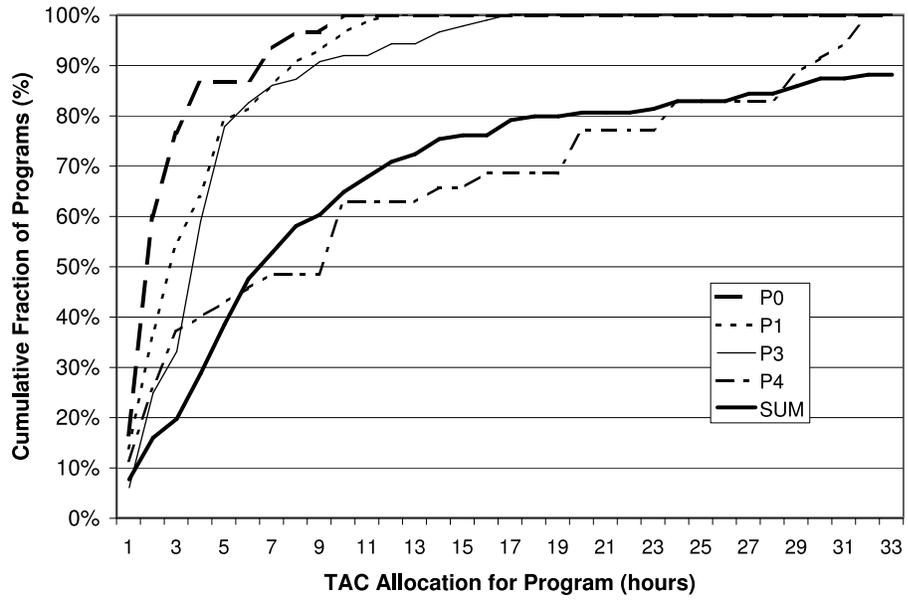}
\caption{Cumulative distribution of TAC allocations in 2006.   Typically
high-priority times are allocated in small increments to a number of programs.
The lowest priority time, P4, is generally allocated in large increments to a
few programs (e.g., 50\% of the P0 programs were allocated 2 hr or less).
\label{fig2}}
\end{figure}

\clearpage

\begin{figure}
\epsscale{0.8}
\plotone{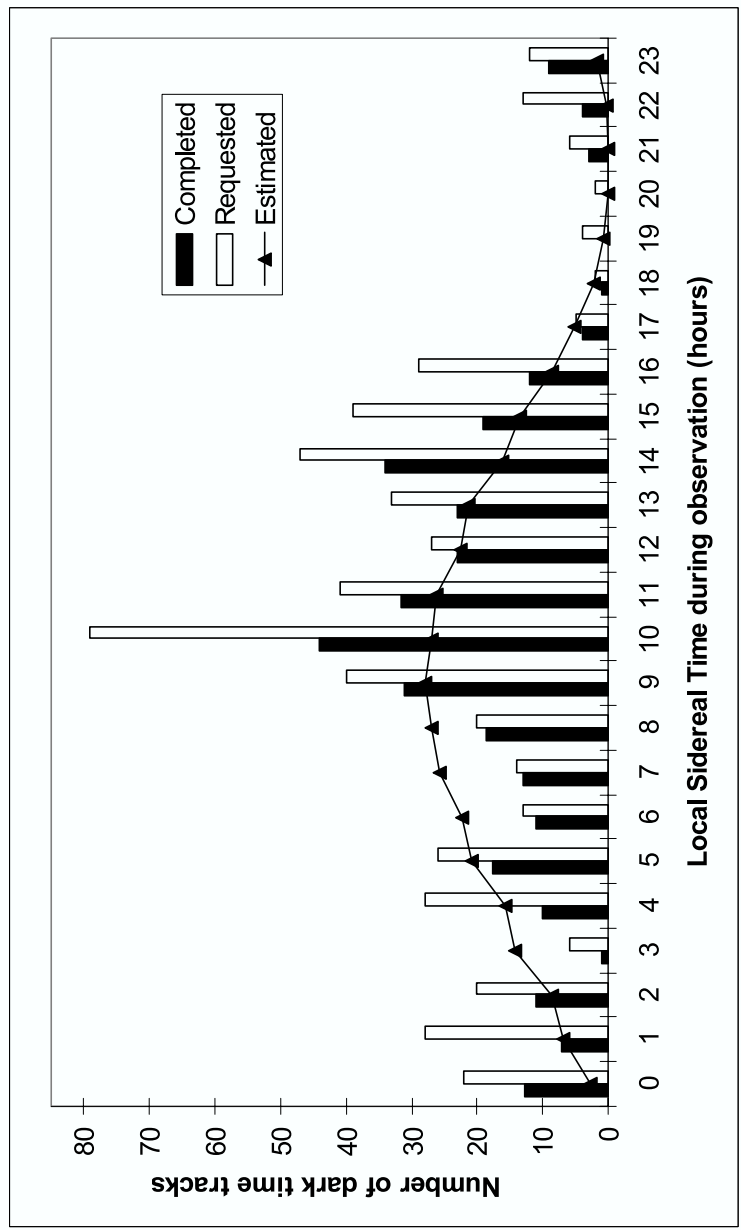}
\caption{Distribution of the completed and requested dark time visits
for priority $1--3$ in the 2006-1 period.   The open histogram represents
the requests, while the filled histogram represents the completed visits.  
The connected triangles gives a model of expected completion rates for this period,
 based on our Web tools.  The completed visits fall above the expected completion
rates because many of the targets can be accessed in two ST windows (east
and west).
\label{fig3}}
\end{figure}

\clearpage

\begin{figure}
\epsscale{0.8}
\plotone{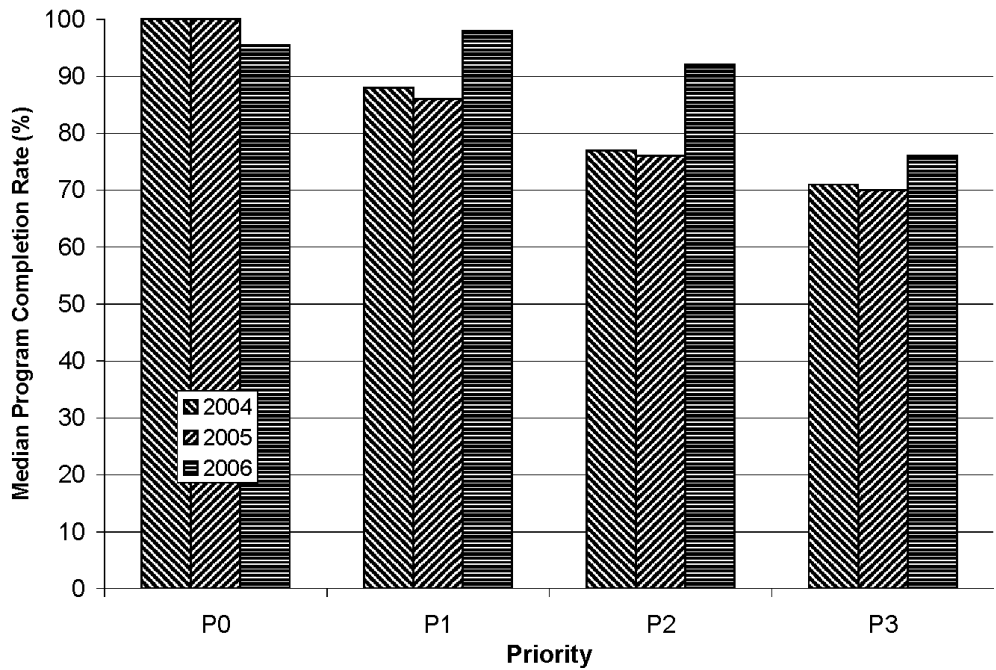}
\caption{
Median individual program completion rates for all programs in
2004, 2005, and 2006.
The completion rate for each program is based on the percentage
of targets or TAC-allocated time successfully completed.  
In 2005, the HET staff adopted the modified-priority algorithm.
In 2006, the User Committee made significant modifications to
this algorithm.
\label{fig4}}
\end{figure}

\clearpage

\begin{figure}
\epsscale{0.8}
\plotone{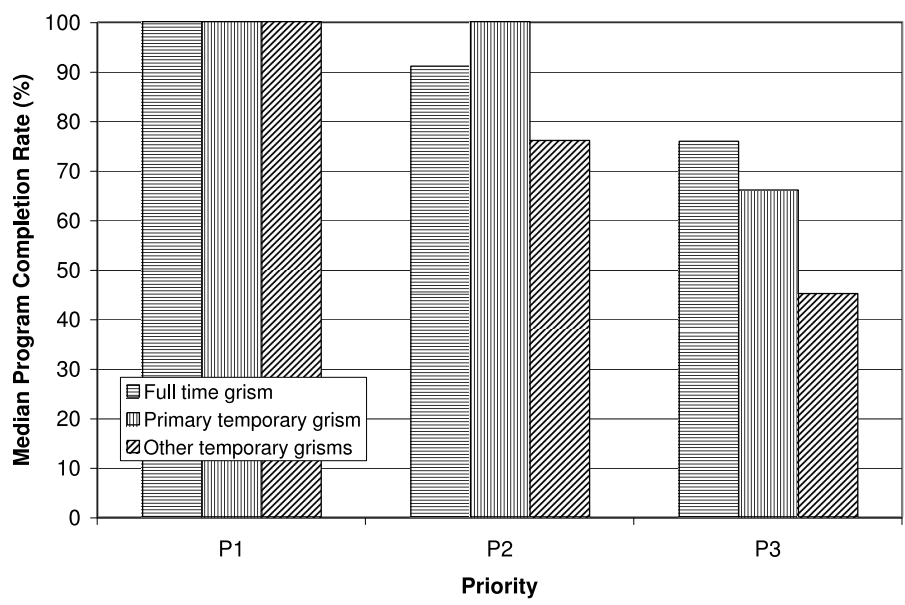}
\caption{
Median individual program completion rates for all programs
in 2005 and 2006 that used either
the permanently mounted grism (LRS\_g1), the most requested
interchangeable grism (LRS\_g2), or the less often requested
interchangeable grisms (LRS\_e2 and LRS\_g3).  
The completion rate for each program is based on the percentage
of targets or TAC-allocated time successfully completed.
\label{fig5}}
\end{figure}

\end{document}